\documentclass[sigconf]{acmart}
\usepackage[ruled,linesnumbered,vlined]{algorithm2e}
\usepackage{multirow}
\usepackage{subfigure}
\usepackage{colortbl}

\AtBeginDocument{%
  \providecommand\BibTeX{{%
    \normalfont B\kern-0.5em{\scshape i\kern-0.25em b}\kern-0.8em\TeX}}}

\copyrightyear{2022}
\acmYear{2022}
\setcopyright{rightsretained} 

\acmConference[CIKM '22]{Proceedings of the 31st ACM International Conference on Information and Knowledge Management}{October 17--21, 2022}{Atlanta, GA, USA} \acmBooktitle{Proceedings of the 31st ACM Int'l Conference on Information and Knowledge Management (CIKM '22), Oct. 17--21, 2022, Atlanta, GA, USA} \acmISBN{978-1-4503-9236-5/22/10} \acmDOI{10.1145/3511808.3557278}

\usepackage{etoolbox}
\makeatletter
\patchcmd{\maketitle}{\@copyrightpermission}{
   \begin{minipage}{0.3\columnwidth}
     \href{https://creativecommons.org/licenses/by/4.0/}{\includegraphics[width=0.90\textwidth]{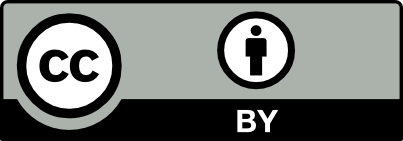}}
   \end{minipage}\hfill
   \begin{minipage}{0.7\columnwidth}
     \href{https://creativecommons.org/licenses/by/4.0/}{This work is licensed under a Creative Commons Attribution International 4.0 License.}
   \end{minipage}
  
   \vspace{5pt}
}{}{}

\makeatother

\begin{document}

\author{Jiaqian Ren}
\affiliation{%
   \institution{Institute of Information Engineering,\\ Chinese Academy of Sciences,}
   \country{Beijing, China}}
\affiliation{%
   \institution{School of Cyber Security,\\ University of Chinese Academy of Sciences,}
   \country{Beijing, China}}
\email{renjiaqian@iie.ac.cn}

\author{Lei Jiang}
\affiliation{%
   \institution{The Second Research Room, \\Institute of Information Engineering,\\ Chinese Academy of Sciences,}
   \country{Beijing, China}}
\email{jianglei@iie.ac.cn}
\authornote{Corresponding authors}

\author{Hao Peng$^*$}
\affiliation{%
   \institution{School of Cyber Science and Technology, \\Beihang University,}
   \country{Beijing, China}}
\email{penghao@buaa.edu.cn}

\author{Lingjuan Lyu}
\affiliation{%
   \institution{Sony AI,}
   \country{Tokyo, 108-0075, Japan}}
\email{Lingjuan.Lv@sony.com}

\author{Zhiwei Liu}
\affiliation{%
   \institution{Salesforce,}
   \country{CA 94301, USA}}
\email{zhiweiliu@salesforce.com}

\author{Chaochao Chen}
\affiliation{%
   \institution{Zhejiang University,}
   \country{Hangzhou 310058, China}}
\email{zjuccc@zju.edu.cn}

\author{Jia Wu}
\affiliation{%
   \institution{Macquarie University}
   \country{Sydney NSW 2109, Australia}}
\email{jia.wu@mq.edu.au}

\author{Xu Bai}
\affiliation{%
   \institution{Institute of Information Engineering,\\ Chinese Academy of Sciences,}
   \country{Beijing, China}}
\email{baixu@iie.ac.cn}

\author{Philip S. Yu}
\affiliation{%
   \institution{University of Illinois Chicago,}
   \country{Chicago, IL 60607, USA}}
\email{ psyu@uic.edu}

\renewcommand{\shortauthors}{Jiaqian Ren, et al.}

\theoremstyle{definition}
\newtheorem{define}{Definition}[]

\title{Cross-Network Social User Embedding with Hybrid Differential Privacy Guarantees}

\begin{abstract}
Integrating multiple online social networks (OSNs) has important implications for many downstream social mining tasks, such as user preference modelling, recommendation, and link prediction. However, it is unfortunately accompanied by growing privacy concerns about leaking sensitive user information. How to fully utilize the data from different online social networks while preserving user privacy remains largely unsolved. To this end, we propose a Cross-network Social User Embedding framework, namely DP-CroSUE, to learn the comprehensive representations of users in a privacy-preserving way. We jointly consider information from partially aligned social networks with differential privacy guarantees. In particular, for each heterogeneous social network, we first introduce a hybrid differential privacy notion to capture the variation of privacy expectations for heterogeneous data types. Next, to find user linkages across social networks, we make unsupervised user embedding-based alignment in which the user embeddings are achieved by the heterogeneous network embedding technology. To further enhance user embeddings, a novel cross-network GCN embedding model is designed to transfer knowledge across networks through those aligned users. Extensive experiments on three real-world datasets demonstrate that our approach makes a significant improvement on user interest prediction tasks as well as defending user attribute inference attacks from embedding.  
\end{abstract}


\maketitle
\section{Introduction}


Social media-based user embedding plays an important role in user representation, user analysis and many downstream applications. Nowadays, to incorporate more information and get enhanced user embeddings, new technologies~\cite{zhang2013predicting,yan2013friend,yan2016unified,zhang2017bl,lin2019cross,cui2021exploiting} which fuse and mine multiple social networks together show promising trends.
However, this trend is now challenged by serious privacy concerns. 
Authors in \cite{kats2018many} report that more than 80\% US Internet users were worried about the usage of their personal data. 
Meanwhile, more rigorous regulations like EU's GDPR\footnote{\url{https://gdpr-info.eu}} are enacted to 
regulate the usage of personal information. For example, companies cannot share a user's data without his/her consent. Henceforth, raw social networks which encode individual's sensitive information (e.g., friendship, gender, occupation) should not be disclosed to others directly. 
This paper initiates the study on privacy-preserving social user embeddings across multiple online social networks (OSNs). 


Nevertheless, the real-world scenarios are complex because OSNs contain different types of information, which are formulated as heterogeneous social networks (HSNs).
To protect privacy, existing works use 1) various anonymization techniques and 2) differential privacy (DP) mechanisms. 
Since anonymization techniques, including $k$-anonymity~\cite{sweeney2002k}, $l$-diversity~\cite{machanavajjhala2007diversity}, $t$-closeness~\cite{li2007t}, etc, may be vulnerable to deanonymization attacks~\cite{narayanan2009anonymizing} and lacking a rigorous theoretical guarantee, we adopt DP in our work.
As a mathematically rigorous privacy-preserving framework, DP has been widely used to release social graphs~\cite{sala2011sharing,wang2013preserving,lu2014exponential,xiao2014differentially,nguyen2016network,liu2020local,yang2020secure,jorgensen2016publishing,ji2019differentially,wang2021differentially, peng2021differentially}. 
Among them, the standard techniques~\cite{sala2011sharing,wang2013preserving,lu2014exponential,xiao2014differentially,nguyen2016network,liu2020local,yang2020secure} only consider to sanitize graph structure and ignore vertex attributes. 
The follow-up works~\cite{ji2019differentially,jorgensen2016publishing,wang2021differentially} make up for this deficiency and study releasing attributed graphs. However, they are limited to homogeneous graphs. Considering real-world social networks contain multiple types of information, and different types of information have different privacy protection expectations, how to publish HSNs with proper differential privacy guarantees remains an open challenge. In this work, we propose a hybrid DP mechanism to handle each kind of data independently.

Suppose we have multiple protected heterogeneous social networks under differential privacy protection, another challenging problem is how to integrate them together to generate more comprehensive user representations. 
Obviously, two key issues in network fusion are user linkage and cross-network information transferring. 
Prior user linkage methods can be divided into 1) user attribute-based ones and 2) embedding-based ones. Attribute-based methods such as~\cite{andreou2017identity,buraya2017towards,chen2012more,yao2021differential} aim to link the same user across different OSNs through the comparison of real user information.
However, those perturbed social networks no longer contain accurate information. Thus, attribute-based methods will fail.
Embedding-based alignment methods~\cite{liu2016aligning, man2016predict, lample2018word, chu2019cross, liang2021unsupervised} have gained lots of attention in recent years. 
It is worth noting that those perturbed networks still preserve important characteristics of the original ones. 
Therefore, if embedding-based methods are trained to capture essential characteristics~\cite{ma2022deep}, they are still able to learn important semantics and achieve alignment. As such, we adopt embedding-based methods in our framework to find user linkages. 
To further leverage multi-source data for improving cross-network analysis, a series of works ~\cite{zhang2013predicting,yan2013friend,yan2016unified,zhang2017bl,lin2019cross,liu2021federated, xu2021privacy,che2022federated,chen2020survey,yang2021consisrec} have made great success in applications such as user profile modelling, social recommendation and so on. 
However, as mentioned before, most of them have 
ignored the privacy leakage problem. 
To sum up, how to support social network integration with proper user privacy protection is an important yet under-explored problem.  

To achieve privacy-preserving social network integration for improving social user embeddings, we propose a novel framework, called DP-CroSUE. 
Particularly, we perturb those heterogeneous social networks before data exchange to protect users' privacy. 
There are various sensitive data types in the social network, including graph topological data (user friendship), multi-dimensional numerical and categorical data (user attributes like age and gender), and text data (the posts), where each data has different privacy expectations.
We introduce a hybrid differential privacy mechanism to generate proper perturbations. Our mechanism ensures data utility while preserving necessary privacy. To find user linkages across networks for further integration, we apply generative adversarial networks (GANs) to conduct unsupervised user alignment. Finally, a novel cross-network GCN embedding model including inter-graph propagation and hierarchy intra-graph propagation is proposed to transfer information both inside and across networks.
We evaluate DP-CroSUE on three real-world social network platforms considering the embedding usefulness for user interest prediction tasks and its ability to resist two user attribute inference attacks: gender inference and occupation inference. Noted that the DP-CroSUE framework can also be generalized to other tasks like recommendation. We mainly focus on user interest prediction tasks for illustration purpose in this paper. 
Experimental results indicate that DP-CroSUE makes a good balance in both user feature utility and user privacy protection. 
The source code and data are available at GitHub\footnote{https://github.com/RingBDStack/DP-CroSUE}.

Our contributions are three-fold: 1) We propose DP-CroSUE, the first attempt to integrate multiple HSNs for comprehensive social user embeddings with a hybrid differential privacy guarantee. Our approach demonstrates a competitive trade-off between user feature utility and user privacy protection. 2) We introduce a hybrid differential privacy mechanism capturing the variation of privacy expectations for heterogeneous social graphs. 3) We propose a novel cross-network GCN embedding model including inter-graph propagation and hierarchy intra-graph propagation to transfer information both inside and across social networks to make complete information integration.





\section{Related Work}

\textbf{Differential Privacy.}
DP~\cite{dwork2006calibrating} has been widely used for privacy-preserving statistical analysis. The intuition behind it is to randomise the output to ensure that the presence of any individual in the input has a negligible impact on the probability of any particular output. Earlier DP mechanisms such as Laplace mechanism and Exponential mechanism are proposed to protect single numerical~\cite{dwork2006calibrating,ding2017collecting} and categorical~\cite{erlingsson2014rappor,wang2017locally} data. Whereas recently, mechanisms have been developed for different data types and domains~\cite{zhou2020vertically,truex2020ldp,feyisetan2019leveraging,feyisetan2020privacy,lyu2020differentially,lyu2020towards}. For example, authors in \cite{truex2020ldp} extend original methods to handle multi-dimensional data. In the NLP domain, a few works directly inject high-dimensional DP noise into text representations~\cite{feyisetan2019leveraging,feyisetan2020privacy,lyu2020differentially,lyu2020towards}. However, due to “the curse of dimensionality”, 
they fail to strike a nice privacy-utility balance. 
Authors in \cite{ACL21YueDu21} solve this problem by sampling a close word substitute to ensure utility. 
In the social network domain, a series of works~\cite{sala2011sharing,wang2013preserving,lu2014exponential,xiao2014differentially,yang2020secure} have been proposed to protect graph structure. They focus on edge-DP and protect graph topologies only. The follow-up works~\cite{jorgensen2016publishing,ji2019differentially} cover this shortage by taking users’ attributes into account. However, they only consider the homogeneous networks. How to perturb heterogeneous social networks is not covered in the literature.

\textbf{User Linkage and Social Network Integration.}
User linkage \cite{narayanan2010myths}, also known as social network alignment, has been an important issue to make further utilization of social network data. Existing works~\cite{andreou2017identity,buraya2017towards,chen2012more} make alignments by carefully comparing the user attributes. These solutions are now challenged by privacy concerns about the disclosure of sensitive user attributes. Recently, authors in \cite{yao2021differential} make the first attempt to study privacy-preserving user linkage across multiple OSNs. However, it still needs some user "volunteers" whose linkages are known in advance. In recent years, embedding-based alignment methods~\cite{liu2016aligning, man2016predict, lample2018word, chu2019cross, liang2021unsupervised} have achieved great success in finding anchor users across two or more social networks, which gives opportunities to make social network integration. Meanwhile, the cross-network results have been proved to enhance various social network applications. For example, some works~\cite{cao2017joint,lim2015mytweet,jiang2016little,ren2020banana} fuse different social networks to provide a better understanding of users’ interests and behaviours. 

\section{Terminology definition and problem formulation}

\subsection{Definitions}\label{Sec_definition}
Differential Privacy~\cite{dwork2006calibrating} has emerged as a strong privacy definition for statistical data release with the intuition that a randomized algorithm behaves similarly on neighbouring datasets.

\begin{definition}\label{DP}
($\epsilon$-DP~\cite{duchi2013local}) \textit{For any $\epsilon > 0$, a randomized mechanism $M$ satisfies $\epsilon$-DP if for any two inputs $x, x'$ in the domain of $M$, and for any output $y$ of $M$, we have:}
\begin{equation}
\frac{\text{Pr}\{M(x)=y\}}{\text{Pr}\{M(x')=y\}} \leq \exp(\epsilon),
\end{equation}
where $Pr\{\cdot\}$ represents probability, $\epsilon$ corresponds to privacy budget. Smaller $\epsilon$ asserts a better privacy protection but lower data utility.
\end{definition}
 Noted that early DP is oriented toward structured data. For the protection of graph data, the notion of $\epsilon$-edge-DP is proposed.

\begin{definition}\label{Edge-DP}
($\epsilon$-Edge-DP~\cite{blocki2012johnson}) \textit{For any $\epsilon > 0$, a randomized mechanism $M$ satisfies $\epsilon$-Edge-DP if for any two neighbouring graphs $G_1,G_2\in \mathcal{G}$, which differ by at most one edge, and for any output $S$ of range($M$), we have:}
\begin{equation}
\frac{\text{Pr}\{M(G_1)=S\}}{\text{Pr}\{M(G_2)=S\}} \leq \exp(\epsilon).
\end{equation}
\end{definition}
Being a very strong privacy notion, $\epsilon$-DP, however, is unsuitable to protect text privacy with utility~\cite{ACL21YueDu21}. Because under $\epsilon$-DP protection, a word will be transformed to any other words with equal probabilities, no matter how unrelated they are. Thus a sanitized token may not capture the semantics. The relaxed notion of Metric DP (MDP) can address this problem.

\begin{definition}\label{MDP}
(MDP~\cite{alvim2018local}) \textit{Given $\epsilon > 0$ and a distance metric $d$, a randomized mechanism $M$ satisfies MDP if for two inputs $x, x'$ in the domain of $M$, and for any output $y$ of $M$, we have:} \begin{equation}
\frac{Pr\{M(x)=y\}}{Pr\{M(x')=y\}} \leq \exp(\epsilon\cdot d(x,x')).
\end{equation}
\end{definition}
For MDP, the indistinguishability of output distributions is further controlled by the corresponding distance between the inputs, and the metric $d$ needs to be defined according to the specific application. 
Considering different characteristics and privacy expectations of heterogeneous social networks, we introduce a novel hybrid-DP 
notion which preserves various graph properties through carefully designing the injected noise.
\begin{definition}\label{HDP}
(Hybrid-DP) \textit{Given $\epsilon_a > 0$ for attribute feature, $\epsilon_g > 0$ for graph edge, $\epsilon_t > 0$ for textual data, and a distance metric $d$, suppose there are two neighbouring heterogeneous social graphs $G$ and $G'$ which differ in one user node's attribute vector, the presence of a single edge and one post,
a randomized mechanism $M$ satisfies hybrid-DP if for any output $O$ of $M$, we have:} \begin{equation}
\frac{Pr\{M(G)=O\}}{Pr\{M(G')=O\}} \leq exp(\epsilon_a + \epsilon_g + \epsilon_t\cdot d(G_t,G_t')),
\end{equation}
where $G_t$ and $G'_t$ represent the textual data (posts) in the graphs.
\end{definition}

\subsection{Problem Formulation}\label{Sec_problem}
\begin{figure*}[t]
    \centering
    \includegraphics[width=0.84\textwidth]{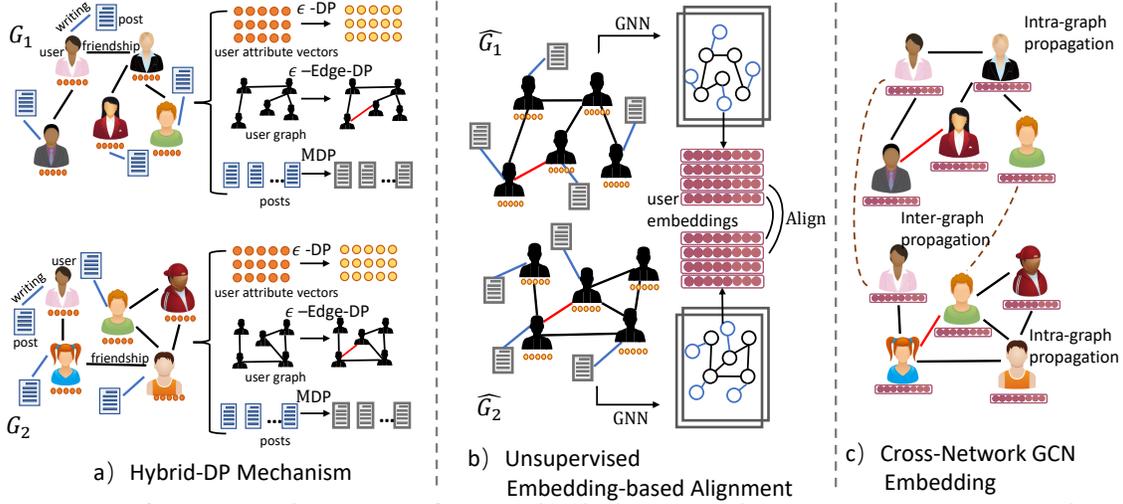}
    \vspace{-6mm}
    \caption{
    An overview of our proposed DP-CroSUE framework, which contains three important steps. 
    First, to share OSNs with each other, each raw heterogeneous social network is perturbed by the hybrid differential privacy mechanism (Section~\ref{Sec_HDPmechanism}).  
    Second, we utilize heterogeneous network embedding technology to get initial user embeddings and make unsupervised network alignment to find anchor users (Section~\ref{Sec_align}). 
    Finally, a novel cross-network GCN embedding model is proposed to get enhanced social user representations by integrating knowledge across networks (Section~\ref{Sec_crossEMB}).
    }
    \label{Fig_model}
    \vspace{-2mm}
\end{figure*}

Generally, each online social network can be represented as a heterogeneous graph. In particular, as shown in Figure~\ref{Fig_model}, we map every single network to a heterogeneous network $G=(V,E,R)$ containing two types of nodes: (i) users $v_u$; (ii) textual posts $v_t$ written by users, and two types of edge relationships: (i) friendship $r_f$ (user-user) and (ii) writing $r_w$ (user-post). The user-type nodes have their respective multi-dimensional user attribute features $\mathbf{X}$. 

The problem of DP-CroSUE is to obtain more comprehensive representations of social users by combining multiple HSNs together without raw data leakage. Formally, this problem can be divided as follows: \textbf{1) Heterogeneous social graph protection with hybrid differential privacy guarantees}: Given a social network company's data which can be represented as a heterogeneous network $G$, we need to design a hybrid randomized mechanism $M_h$: $M_h(G)\to \hat{G}$ by fully considering the different characteristics and privacy expectations of different data types. That is to say, for each data type, we should adopt the proper DP notion and set a proper privacy budget. \textbf{2) Cross-network information transferring}:
 Given two protected social networks $\hat{G}_1$ and $\hat{G}_2$, to transfer knowledge across these two networks, a set of anchor users need to be found. Next, with those anchor pairs working as a "bridge", we need to design a transferring architecture which propagates information across and within these two networks effectively.


\section{Model}
In this section, we describe DP-CroSUE in detail. The whole framework is shown in Figure~\ref{Fig_model}.

\subsection{Hybrid-DP Mechanism}\label{Sec_HDPmechanism}
As mentioned in Section \ref{Sec_problem}, we consider privacy protection of three types of sensitive information in heterogeneous social networks - user attribute features, user friendship relationships and posts written by users. They correspond to three data formats respectively: multidimensional numerical and categorical data, graph edge data and textual data. 
Given the different characteristics of different data formats and users' heterogeneous privacy expectations, treating all the data types as equally sensitive will add too much unneeded noise and sacriﬁce utility. 
For example, the strict $\epsilon-DP$ notion applied in text data will totally change the original semantics and result in low utility. Besides, the contributions of a user's attribute information, the numerous posts he/she has posted and the friendship relations to the final prediction tasks and information leakage differ a lot. Thus, a hybrid differential privacy mechanism which carefully designs the injected noise is highly needed. Specifically, we adopt $\epsilon$-DP, $\epsilon$-Edge-DP and MDP for attribute data, graph edge data and textual data respectively. Meanwhile, we also give insights into allocating proper privacy budgets. 

We first introduce the three strategies for different data formats.
To protect user attributes, 
we directly inject noise to the user attribute vector containing both numerical and categorical features. The extended PM algorithm proposed in \cite{zhang2021graph} is adopted. According to the proof in \cite{zhang2021graph}, it satisfies $\epsilon$-DP. For the protection of user friendship relations, we extract the homogeneous user graph and enforce edge-DP to it. Specifically, we adopt TmF \cite{nguyen2015differentially} algorithm, which first computes a new, noisy number of graph edges, then utilizes a filter to decide whether the original edge should be preserved or not. As proved in \cite{nguyen2015differentially}, TmF satisfies $\epsilon$-Edge-DP.



To protect text with utility, we develop a sanitation mechanism with a MDP guarantee. First, we inject each word in the corpus $C$ to an embedding. We denote the injection function $\phi$, which can be any of the well-known word embedding algorithms (e.g., Word2Vec  \cite{mikolov2013efficient}, GloVe \cite{pennington2014glove}, or FastText \cite{bojanowski2017enriching}). Here we select Word2Vec\cite{mikolov2013efficient}. For any two words $x$ and $x'$, we define their distance $d(x,x')=d_{euc}(\phi(x),\phi(x'))$, where $d_{euc}$ represents the Euclidean distance. To achieve MDP, for each word $x$, we run the randomized mechanism $M$ to sample a sanitized $y$ with probability:
\begin{equation}\label{M-MDP}
Pr\{M(x)=y\}=\frac{exp(-\frac{1}{2}\epsilon\cdot d_{euc}(\phi(x),\phi(y)))}{\sum_{y'\in C}exp(-\frac{1}{2}\epsilon\cdot d_{euc}(\phi(x),\phi(y')))}.
\end{equation} 

\begin{theorem}
\textit{Given $\epsilon > 0$ and a distance metric $d_{euc}$, the randomized mechanism $M$ depicted in Eq.~\ref{M-MDP} satisfies MDP.}
\label{theorem_text}
\end{theorem}

Proof of Theorem~\ref{theorem_text}. Consider a sentence $D$ only having one word <$x$>, another sentence $D'$ which is <$x'$> ($x'\not= x$), and a possible output $y$. We set $C_x = ({\sum_{y'\in C}exp(-\frac{1}{2}\epsilon\cdot d_{euc}(\phi(x),\phi(y')))})^{-1}$.
\begin{equation}
\begin{split} 
\frac{Pr\{M(x)=y\}}{Pr\{M(x')=y\}}
=\frac{C_x\cdot exp(-\frac{1}{2}\epsilon\cdot d_{euc}(\phi(x),\phi(y)))}{C_{x'}\cdot exp(-\frac{1}{2}\epsilon\cdot d_{euc}(\phi(x'),\phi(y)))}\\
=\frac{C_x}{C_{x'}}\cdot exp(\frac{1}{2}\epsilon\cdot [d(x',y)-d(x,y)]) \\
\leq\frac{C_x}{C_{x'}}\cdot exp(\frac{1}{2}\epsilon\cdot d(x,x'))\\
\leq\frac{\sum_{y'\in C}exp(-\frac{1}{2}\epsilon\cdot d_{euc}(\phi(x),\phi(y')))}{\sum_{y'\in C}exp(-\frac{1}{2}\epsilon\cdot d_{euc}(\phi(x'),\phi(y')))}\cdot exp(\frac{1}{2}\epsilon \cdot d(x,x'))\\
\leq exp(\frac{1}{2}\epsilon \cdot d(x,x'))\cdot exp(\frac{1}{2}\epsilon \cdot d(x,x'))=exp(\epsilon\cdot d(x,x')).
\end{split}
\end{equation}

We now analyze the overall differential privacy guarantee by combining all the above three information perturbations. 
We have the following theorem:
\begin{theorem}\label{theorem_HDP}
Assume that we independently adopt the three perturbation algorithms described above and the attribute feature, graph edge, and textual data satisfy $\epsilon_a$-DP, $\epsilon_g$-Edge-DP, and $\epsilon_t$-MDP, respectively. This hybrid perturbation mechanism satisfies our hybrid-DP notion defined in Definition \ref{HDP}.
\end{theorem}

Proof of Theorem~\ref{theorem_HDP}. We assume the three kinds of data in the heterogeneous graph $G$ are independent and denote the three perturbation mechanisms for user attributes, graph structure and user posts as $M_a$, $M_g$, $M_t$, respectively. Meanwhile, we utilize $G_a$, $G_g$ and $G_t$ to represent the attribute features, graph structure and posts alone. $G_a'$, $G_g'$ and $G_t'$ are the corresponding neighbouring datasets for each data type. Noted that the hybrid-DP mechanism $M_h$ is the combination of the three randomized mechanisms mentioned above and $G'$ denotes the whole perturbed heterogeneous social network. Since the attribute part, graph edge part and textual part satisfies $\epsilon$-DP, $\epsilon$-Edge-DP and MDP, respectively, we have:

\begin{equation}
\begin{split} 
\frac{\text{Pr}\{M_h(G)=y\}}{\text{Pr}\{M_h(G')=y\}} = \frac{\text{Pr}\{M_a(G_a)=y_a, M_g(G_g)=y_g, M_t(G_t)=y_t\}}{\text{Pr}\{M_a(G_a')=y_a, M_g(G_g')=y_g, M_t(G_t')=y_t\}}\\
=\frac{\text{Pr}\{M_a(G_a)=y_a\}}{\text{Pr}\{M_a(G_a')=y_a\}}\cdot\frac{\text{Pr}\{M_g(G_g)=y_g\}}{\text{Pr}\{M_g(G_g')=y_g\}}\cdot\frac{\text{Pr}\{M_t(G_t)=y_t\}}{\text{Pr}\{M_t(G_t')=y_t\}}\\
\leq \exp(\epsilon_a)\cdot\exp(\epsilon_g)\cdot\exp(\epsilon_t\cdot d(G_t,G_t'))\\
=\exp(\epsilon_a + \epsilon_g + \epsilon_t\cdot d(G_t,G_t') ).
\end{split}
\end{equation}

Note that these three perturbation strategies are designed according to the characteristics of different data formats.
Additionally, how to set privacy budgets ($\epsilon_a$, $\epsilon_g$ and $\epsilon_t$) to achieve proper privacy levels as well as utility for heterogeneous graph properties is also challenging. 
We solve this by referring to the Task-relevance to Message-inference Ratio (TMR). The intuition behind is that less noise being injected into the extracted data type which is more relevant to the target task will bring out more utility. Meanwhile, more noise should be injected into those data types causing accurate personal message inference to satisfy privacy.
In this work, we leverage the precision score obtained through a single piece of information in the interest prediction task to measure task relevance and use the sum of precision scores in inference attacks as the message-inference value.
Since larger privacy budget means less injected noise, we set larger $\epsilon$ for data with high TMR values. More details can be seen in Section~\ref{baseline}.

\begin{table*}
    \caption{Statistics of datasets used in the experiments. }
    \centering
    \renewcommand\arraystretch{1.05}
    \setlength{\tabcolsep}{4mm}
    \begin{tabular}{c|c|c|c|c|c|c}
    \toprule
    \multirow{2}{*}{Dataset} & \multirow{2}{*}{Network} &\multicolumn{3}{c|}{ Nodes} &\multicolumn{2}{c}{Relationships}  \\
    \cline{3-7}
    &&\#(Users)&\#(Anchor Users)&\#(Posts)&\#(friendship)&\#(write)\\
    \hline
    \multirow{2}{*}{Foursquare-Twitter}& Foursquare &5392 &\multirow{2}{*}{3388} &48756 &76972 &48756 \\
    \cline{2-3} \cline{5-6}
    & Twitter &5223 & &615,515 &164,920  &615,515 \\
    \hline
    \multirow{2}{*}{Weibo}& Sub-weibo1 &8117 &\multirow{2}{*}{2969} &158,823 &12000 &158,823\\
    \cline{2-3} \cline{5-6}
    & Sub-weibo2 &8539 &\multirow{2}{*}{} &153,741 &12000 &153,741\\
    \hline
\end{tabular}
\label{table_statistic}
\end{table*}

\subsection{Embedding-based Social User Alignment}
\label{Sec_align}
We leverage embedding-based alignment methods to find anchor user linkages, which is divided into two procedures: 1) heterogeneous social network embedding learning and 2) unsupervised user linkage prediction. 

To fully encode all kinds of information in the perturbed heterogeneous social network $\hat{G}=(V,\hat{E},R)$, we adopt relation-specific transformations and the propagation function is:
\begin{equation}
\mathbf{h}_i^{(l+1)}=\sigma\begin{pmatrix}
\sum_{r\in R}\sum_{j\in\mathcal{N}_i^r}\frac{1}{c_{i,r}}\mathbf{W}_r^{(l)}\mathbf{h}_j^{(l)}+\mathbf{W}_0^{(l)}\mathbf{h}_j^{(l)}
\end{pmatrix},
\end{equation}
where $\mathbf{h}_i^{l+1}$ is the updated representation in the $(l+1)$-th layer. $\mathcal{N}_i^r$ represents the set of one-hop neighbor indices of node $i$ under relation $r\in R$ and $c_{(i,r)}=|\mathcal{N}_i^r|$. 
Suppose the final user embedding is $\mathbf{z}_{v_u}$ ($v_u\in \mathcal{V}_u$), the loss function can be expressed as:
\begin{equation}
L_\mathcal{\hat{G}} = -\log\begin{pmatrix}\sigma(\mathbf{z}_{v_u}^{\top}\mathbf{z}_{v_u'})\end{pmatrix}-Q\cdot \mathbb{E}_{v_n\sim P_n(V_u)}\log\begin{pmatrix}\sigma(\mathbf{z}_{v_u}^{\top}\mathbf{z}_{v_n}) \end{pmatrix},
\label{Eq_graphloss}
\end{equation}
where $v_u'$ is the one-hop neighbours of $v_u$. $P_n(V_u)$ defines negative sampling distribution of user nodes. $Q$ is the number of negative samples.

Next, we leverage the obtained social user embeddings $\mathbf{Z}_{u1}$ and $\mathbf{Z}_{u2}$ to make the alignment. As the spaces of $\mathbf{Z}_{u1}$ and $\mathbf{Z}_{u2}$ are learnt independently, we need to learn the matrix $\mathbf{W}$ such that $\mathbf{W}=\mathop{argmin}||\mathbf{WZ}_{u1}-\mathbf{Z}_{u2}||$ to reconcile them. Here the source network is $\hat{G}_1$ and the target one is $\hat{G}_2$. We get $\mathbf{W}$ by Generative Adversarial Networks (GANs), where the generator learns the transformation matrix $\mathbf{W}$, ensuring that the transformed $\mathbf{WZ}_{u1}$ approximates $\mathbf{Z}_{u2}$ as closely as possible. The discriminator tries to classify whether the embeddings are real $\mathbf{Z}_{u2}$ or those transformed ones. 
When $\mathbf{W}$ is trained well, we utilize the similarity measure CSLS proposed in \cite{lample2018word} to find equivalent users.

\begin{figure}[t]
    \centering
    \includegraphics[width=0.46\textwidth]{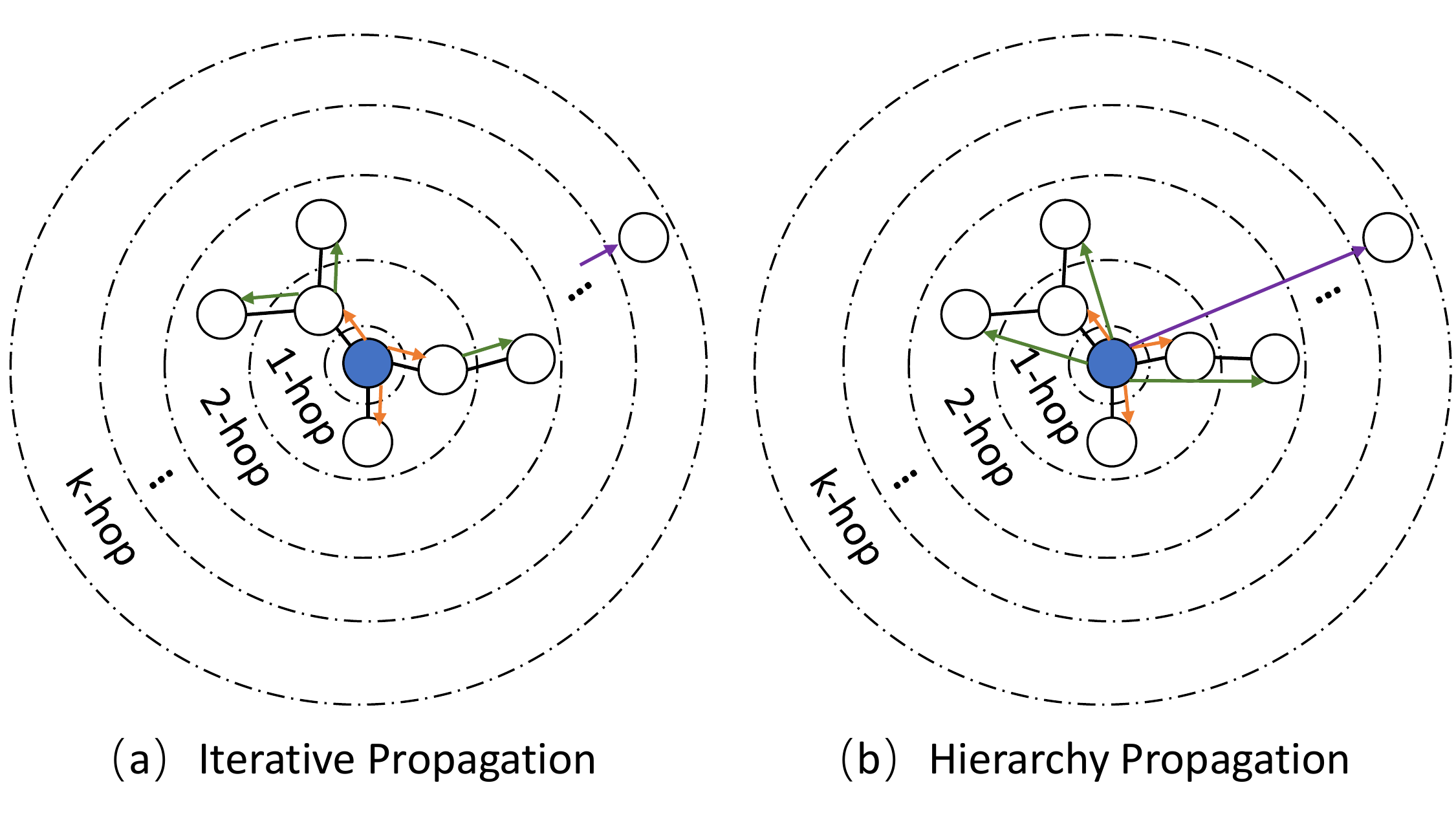}
    \vspace{-6mm}
    \caption{
    The propagation processes of the iterative architecture in GCN and the hierarchy architecture in DP-CroSUE. The node in blue represents the anchor user.
    }
    \label{Fig_prop}
    \vspace{-2mm}
\end{figure}

 \begin{table}
    \caption{The TMR values of attribute, friendship and posts data.}
    \centering
    \renewcommand\arraystretch{1.05}
    \setlength{\tabcolsep}{1mm}
    \begin{tabular}{c|c|c|c|c}
    \toprule
    data &Foursquare &Twitter &Sub-Weibo1 &Sub-Weibo2 \\
    \hline
    attribute&0.287&0.274&0.118&0.144 \\ 
    \hline
    friendship&0.743&0.665&0.225&0.249\\ 
    \hline
    posts&0.645&0.537& 0.207&0.222          \\
    \toprule
\end{tabular}
\label{budget}
\end{table}
\subsection{Cross-network GCN Embedding Model}
\label{Sec_crossEMB}
After finding anchor users, we further integrate the partially aligned homogeneous user networks $\hat{G}_{u1}=(V_{u1},\hat{E}_{u1})$ and $\hat{G}_{u2}=(V_{u2},\hat{E}_{u2})$. To integrate them together, we propose a novel cross-network GCN embedding model, which is composed by an inter-graph propagation layer and a hierarchy intra-graph propagation layer.

$\bullet$ \textbf{inter-graph propagation layer:} We first update those aligned user nodes in both networks by fusing their information together. Considering the existing network disparity of $\hat{G}_{u1}$ and $\hat{G}_{u2}$, we leverage transformation matrices $\mathbf{W}_{12}$ and $\mathbf{W}_{21}$ to project embeddings into the other network space. For example, $\mathbf{W}_{12}$ is used to project users of $\hat{G}_{u1}$ into the space of $\hat{G}_{u2}$. 
Suppose two social users $v_{u1}^{anchor}$ and $v_{u2}^{anchor}$ are aligned, the representations of them after inter-graph propagation are:
\begin{equation}
	\begin{split} 
    \mathbf{z}_{u1}'^{\text{anchor}}=\sigma\begin{pmatrix}\mathbf{z}_{u1}^{anchor}+\mathbf{W}_{21}\mathbf{z}_{u2}^{anchor}\end{pmatrix},\\
    \mathbf{z}_{u2}'^{anchor}=\sigma\begin{pmatrix}\mathbf{z}_{u2}^{anchor}+\mathbf{W}_{12}\mathbf{z}_{u1}^{anchor}\end{pmatrix},
    \end{split}
\end{equation}
where $\sigma$ denotes the activation function.

$\bullet$ \textbf{hierarchy intra-graph propagation layer:}
We set the two user embeddings $Z_{u1}'$ and $Z_{u2}'$ with those aligned users getting updated through inter-graph propagation, while others are still the initial representations learned by GNN model. Now those anchor users' representations contain knowledge from both networks. Considering the number of anchor users is limited, propagating the cross-graph information from those aligned users to the entire graph needs more hops. Considering the over-fitting and over-smoothing problems of GCN~\cite{liu2020towards}, we leverage a hierarchy intra-graph propagation layer which emphasizes the role of anchor users and directly transfers information to nodes within $k$-hop range. 
Suppose $\mathbf{\hat A}_{u1}$ and $\mathbf{\hat A}_{u2}$ are the normalized adjacency matrices of the two user networks, and $\mathbf{D}_{u1}^{anchor}$ and $\mathbf{D}_{u2}^{anchor}$ are the diagonal matrices with the positions of anchors being set to one while others are zeroes, the propagation function is:
\begin{equation}
	\begin{split} 
	\mathbf{H}_{u1}^l=(\alpha\cdot\mathbf{D}_{u1}^{anchor}+\mathbf{\hat A}_{u1}^l)\cdot\mathbf{Z}_{u1}',(l=1,2,...,k),\\
	\mathbf{H}_{u2}^l=(\alpha\cdot\mathbf{D}_{u2}^{anchor}+\mathbf{\hat A}_{u2}^l)\cdot\mathbf{Z}_{u2}',(l=1,2,...,k),\\
	\mathbf{O}_{u1} = \sigma\begin{pmatrix}\mathbf{W}_{u1}\cdot\text{stack}(\mathbf{Z}_{u1}',\mathbf{H}_{u1}^1,...,\mathbf{H}_{u1}^k)\end{pmatrix},\\
	\mathbf{O}_{u2} = \sigma\begin{pmatrix}\mathbf{W}_{u2}\cdot\text{stack}(\mathbf{Z}_{u2}',\mathbf{H}_{u2}^1,...,\mathbf{H}_{u2}^k)\end{pmatrix}.\\
	\end{split}
\end{equation}

To show the difference between the original iterative propagation layer in GCN and our hierarchy one clearly, we depict their propagation processes in Figure~\ref{Fig_prop}. As shown in Figure~\ref{Fig_prop}(a), for the iterative architecture, it needs $k$ transformations to transfer the carrying knowledge in the anchor user to its $k$-hop neighbours. For example, the forward propagation process of iterative architecture in $\hat{G}_{u1}$ is formulated as $\mathbf{O}^{(u1)}=\sigma\left(\mathbf{\hat A}_{u1}\left(\ldots\left(\sigma\left(\mathbf{\hat A}_{u1} \mathbf{Z}_{u1}' \mathbf{W}_{u1}^{(0)}\right) \ldots\right) \mathbf{W}_{u1}^{(k-1)}\right)\right.$. Obviously, when $k$ becomes large, there will be too many transformation parameters $\mathbf{W}_{u1}^{(i)}$ which will make the network difficult to train. Meanwhile, according to ~\cite{liu2020towards}, if we apply iterative architecture to a connected graph, when $k$ goes to infinity, nodes become indistinguishable. However, as for the hierarchy architecture, each anchor user can be transferred to other users inside $k$-hops directly. It only needs one transformation thus the knowledge of anchor users will not be changed too much when they are transferred to distant users. Meanwhile, the emphasis of anchor users can further help transfer cross-network knowledge. 
By considering the information of multiple hops together, the hierarchy architecture has more capacities in capturing its local information compared to the iterative architecture, which helps make users distinguishable.

$\bullet$ \textbf{Objective Function:} We jointly train these two networks together to integrate them. The loss includes three parts: the graph-based losses $\mathcal{L}_{\hat{G}_{u1}}$ and $\mathcal{L}_{\hat{G}_{u2}}$ computed as per Eq.~\ref{Eq_graphloss}, and the hard alignment regularization which measures the anchor nodes' representation distance in the two networks to regularize the output representation spaces. Overall, the total loss is:
\begin{equation}
	\begin{split} 
    \mathcal{L}_t =\mathcal{L}_{\hat{G}_{u1}}+\mathcal{L}_{\hat{G}_{u2}}+d(\mathbf{O}_{u1}^{anchor}, \mathbf{O}_{u2}^{anchor}).
    \end{split}
    \label{Eq_totalloss}
\end{equation}

\section{Experiments}
In this section, we evaluate DP-CroSUE on both user interest prediction capacities and privacy protection strength. Specifically, we aim to answer the following questions: 
\textbf{Q1}: Compared with undisturbed single network knowledge, can the cross-network results of DP-CroSUE get improved in user interest prediction tasks? \textbf{Q2}: Can DP-CroSUE effectively protect sensitive user attribute information? \textbf{Q3}: How does each part of the perturbation paradigm of different information affect the accuracy of user interest prediction and the privacy-preserving property of DP-CroSUE? \textbf{Q4}: How does the novel cross-network GCN embedding model perform?

\subsection{Datasets}
We select three online social platforms: Foursquare, Twitter and Weibo, and make two partially-aligned dataset pairs. One is the Foursquare-Twitter pair from \cite{zhang2017bl}. For the other, we crawled the Weibo platform and divided it into two sub-networks containing a part of sharing users. We manually labelled users from 10 possible interest categories according to their biographies and posts, including: (1) travel; (2) art; (3) health; (4) food; (5) technology; (6) sports; (7) business; (8) politics; (9) game; (10) fashion. Each user may be classified into multiple interests (multi-label classification).
Table~\ref{table_statistic} shows the detailed statistical information about these datasets. The corresponding user attribute information extracted form the datasets are: \textbf{(1) user screen name} (We decompose the name into sub-string sets and convert them into numerical values using SimHash~\cite{sadowski2007simhash}.), \textbf{(2) number of followers}, \textbf{(3) number of followees}, \textbf{(4) number of posts}, \textbf{(5) gender}, \textbf{(6) occupation}.

\begin{table*}[t]
    \caption{Overall performance for the user interest prediction tasks. Higher precision and F1 score represent better interest prediction performance. (Bold: best; Underline: runner-up.)}
    \centering
    \renewcommand\arraystretch{0.80}
    \setlength{\tabcolsep}{1.5mm}
    \begin{tabular}{c||c|c|c|c||c|c|c|c}
    \toprule
      \multirow{2}{*}{Methods}&\multicolumn{2}{c|}{Foursquare} &\multicolumn{2}{c||}{Twitter} &\multicolumn{2}{c|}{Sub-Weibo1} &\multicolumn{2}{c}{Sub-Weibo2}\\
    \cline{2-9}
    &Precision &Micro-f1 &Precision &Micro-f1 &Precision &Micro-f1 &Precision &Micro-f1 \\
    \hline
    Word2Vec~\cite{mikolov2013efficient} & 0.463±0.013 & 0.455±0.009 &0.463±0.002  &0.450±0.003  &0.226±0.001 &0.238±0.002  &0.231±0.011 &0.236±0.009 \\
    SANTEXT~\cite{ACL21YueDu21} &0.457±0.022 &0.446±0.018 &0.446±0.020 &0.442±0.015 &0.207±0.012 &0.193±0.020 &0.229±0.019 &0.233±0.016\\
    DeepWalk~\cite{perozzi2014deepwalk} &0.453±0.012 &0.447±0.008 &0.413±0.010 &0.409±0.007 &0.215±0.003 &0.176±0.004 &0.247±0.004 &0.243±0.004\\
    SNE~\cite{liao2018attributed} &0.509±0.006 &0.511±0.005 &0.554±0.007 &0.547±0.007 &0.249±0.012 &0.238±0.011 &0.275±0.003 &0.274±0.002\\
    UDMF~\cite{zhang2018user} &0.515±0.003 &0.514±0.003 &0.523±0.007 &0.515±0.006 &0.254±0.004 &0.252±0.005 &0.308±0.003 &0.306±0.002\\

    R-GCN~\cite{schlichtkrull2018modeling}&0.530±0.002 &0.522±0.001  &0.546±0.001 &0.542±0.001 &0.283±0.002 &0.279±0.002 &0.317±0.007 &0.312±0.005\\
    \hline
    DP-R-GCN &0.514±0.001 &0.510±0.002 &0.534±0.004 &0.532±0.003 &0.235±0.001 &0.230±0.002 &0.271±0.006 &0.270±0.005\\
    DP-CroSUE &\underline{0.549±0.006} &\underline{0.541±0.005} &\underline{0.561±0.001} &\underline{0.558±0.001} &\underline{0.303±0.003} &\underline{0.296±0.002} &\underline{0.332±0.003} &\underline{0.334±0.002}\\ 
    CroSUE &\textbf{0.556±0.004} &\textbf{0.551±0.003} &\textbf{0.571±0.002} &\textbf{0.569±0.001} &\textbf{0.332±0.001} &\textbf{0.330±0.001} &\textbf{0.360±0.004} &\textbf{0.361±0.003}\\
    
    \toprule
\end{tabular}
\label{table_interest}
\end{table*}

\subsection{Baseline and Hyperparameter Setting}\label{baseline}
We compare DP-CroSUE with the following methods.

$\bullet$ \textbf{Single-view methods:}
In general, single-view methods can be divided into text-based ones and structure-based ones. We select \textbf{Word2Vec}~\cite{mikolov2013efficient}, \textbf{SANTEXT}~\cite{ACL21YueDu21}, and \textbf{DeepWalk}~\cite{perozzi2014deepwalk} as single-view baselines. Among them, Word2Vec maps a sequence of social media posts into a vector representation. SANTEXT gets text representations under differential privacy. DeepWalk learns the latent representations of users in the homogeneous user network.

$\bullet$ \textbf{Multi-view methods:}
For multi-view methods, we choose \textbf{SNE}~\cite{liao2018attributed}, \textbf{UDMF}~\cite{zhang2018user} and \textbf{R-GCN}~~\cite{schlichtkrull2018modeling} as multi-view baselines. SNE learns user representations by preserving both structural proximity and attribute proximity. UDMF is a novel hybrid DNN-based framework that fuses information across different modalities.
R-GCN aggregates information on HSNs based on different relations.

$\bullet$ \textbf{Variants of DP-CroSUE:}
We utilize \textbf{R-GCN} and \textbf{DP-R-GCN} to observe the performances of single-network user embeddings obtained on the raw heterogeneous social networks and the perturbed ones, respectively. We also create a non-private model called \textbf{CroSUE}, which is a variant of DP-CroSUE deleting the Hybrid DP perturbation mechanism, to show the performance of the cross-network user embeddings without privacy protection.

In DP-CroSUE, to get the perturbed heterogeneous social graph, we set the privacy budget $\epsilon_a=5$ for user attribute feature perturbation, $\epsilon_g=10$ for user graph edge perturbation and $\epsilon_t=7.5$ for textual data perturbation. Our guidance is the TMR values as shown in Table~\ref{budget}. The TMR values of attribute information and textual information are obtained through the corresponding attribute features and sentence embeddings (i.e., averaging word embeddings~\cite{mikolov2013efficient}), respectively, while the TMR of friendship information is obtained through the features learnt by DeepWalk~\cite{perozzi2014deepwalk}.
To get initial user embeddings, we use a two-layer R-GCN. The first layer embedding dimension is 256 and the second embedding dimension is 128. For the cross-network GCN model, we set $k=4$ and $\alpha=2$ in the hierarchy intra-graph propagation layer. The final embedding dimension is 128.

\subsection{Evaluation Metrics}
\label{Sec_evaluation_metric}
The evaluation metrics for user interest prediction and attribute inference attack are as follows:

\textbf{User Interest Prediction.} After getting the social user embeddings, we utilize the decision tree classifier to conduct multi-label user interest classification. We randomly
pick 80\% of data for training, while the rest 20\% is for evaluation. To be more convincing, we report the mean and standard deviation of the results after repeating experiments for 5 times. The specific metrics we leverage to judge the quality are precision and micro-$F1$ score. 

\textbf{Attribute Inference Attack.}
To evaluate all models’ robustness and capabilities in preserving sensitive user attribute information, we quantify the privacy leakage in social user embeddings through two inference attacks: gender inference and occupation inference. For gender inference, it is a binary classification problem. We use Logistic Regression as the attacker to make classification. For the occupation inference attack, we use the decision tree classifier to make classification.
Similarly, we repeat the experiments for 5 times.


\begin{table*}[t]
    \caption{Overall performance for user gender and occupation inference attacks. Lower Precision and F1 scores represent better privacy protection. (Bold: best; Underline: runner-up.)}
    \centering
    \renewcommand\arraystretch{0.8}
    \setlength{\tabcolsep}{0.6mm}
    \begin{tabular}{c|c||c|c|c|c||c|c|c|c}
    \toprule
     \multirow{2}{*}{Attribute}& \multirow{2}{*}{Methods}&\multicolumn{2}{c|}{Foursquare} &\multicolumn{2}{c||}{Twitter} &\multicolumn{2}{c|}{Sub-Weibo1} &\multicolumn{2}{c}{Sub-Weibo2}\\
    \cline{3-10}
    & &Precision &Micro-f1 &Precision &Micro-f1 &Precision &Micro-f1 &Precision &Micro-f1 \\
    \hline
    \multirow{8}{*}{Gender}&Word2Vec~\cite{mikolov2013efficient} & 0.569±0.001 & 0.587±0.001 & 0.705±0.001 & 0.700±0.001 &0.570±0.001  &0.567±0.001  &0.533±0.001  &0.621±0.001 \\
    &SANTEXT~\cite{ACL21YueDu21} &0.559±0.001 &0.578±0.001 &0.644±0.001 &0.639±0.001 &0.560±0.001 &0.555±0.001 &0.517±0.001 &0.614±0.001\\
    &DeepWalk~\cite{perozzi2014deepwalk} &\textbf{0.508±0.001} &\textbf{0.510±0.001} &\textbf{0.501±0.001} &\textbf{0.509±0.002} &\textbf{0.506±0.001} &\textbf{0.517±0.001} &\textbf{0.498±0.001} &\textbf{0.583±0.001}\\
    &SNE~\cite{liao2018attributed} &0.615±0.001 &0.624±0.001 &0.682±0.001 &0.684±0.001 &0.590±0.001 &0.591±0.001 &0.549±0.001 &0.647±0.001\\
    &UDMF~\cite{zhang2018user} &0.995±0.001 &0.995±0.001 &0.999±0.001 &0.999±0.001 &0.999±0.001 &0.999±0.001 &0.998±0.001 &0.998±0.001\\

    &R-GCN~\cite{schlichtkrull2018modeling} &0.791±0.001 &0.794±0.001 &0.738±0.001 &0.739±0.001 &0.727±0.001 &0.727±0.001 &0.684±0.001 &0.727±0.001\\
    \cline{2-10}
    &DP-R-GCN &0.568±0.002 &0.579±0.001  &\underline{0.624±0.001} &\underline{0.627±0.001} &0.538±0.001 &0.537±0.001 &0.518±0.001 &0.617±0.001\\
    &DP-CroSUE &\underline{0.554±0.001} &\underline{0.575±0.001} &0.640±0.001 &0.644±0.001 &\underline{0.536±0.001} &\underline{0.535±0.001} &\underline{0.506±0.001} &\underline{0.601±0.001}\\
    &CroSUE &0.655±0.001 &0.644±0.001 &0.703±0.001 &0.704±0.001 &0.649±0.001 &0.649±0.001 &0.611±0.001 &0.674±0.001\\
    \hline
    \hline
    Attribute&Methods &Precision &Micro-f1 &Precision &Micro-f1 &Precision &Micro-f1 &Precision &Micro-f1\\
    \hline
    \multirow{8}{*}{Occupation}&Word2Vec~\cite{mikolov2013efficient} & 0.149±0.023 & 0.147±0.022 & 0.158±0.007 & 0.154±0.008 & 0.524±0.018 & 0.528±0.017 & 0.508±0.007 & 0.519±0.007\\
    &SANTEXT~\cite{ACL21YueDu21} &0.139±0.012 &0.140±0.018 &0.143±0.010 &0.139±0.007 &0.510±0.013 &0.513±0.012 &0.497±0.004 &0.490±0.014\\
    &DeepWalk~\cite{perozzi2014deepwalk} &\textbf{0.102±0.016} &\textbf{0.098±0.015} &\textbf{0.120±0.003} &\textbf{0.118±0.005} &\textbf{0.450±0.001} &\textbf{0.462±0.001} &0.495±0.023 &0.484±0.028\\
    &SNE~\cite{liao2018attributed} &0.187±0.019 &0.176±0.018 &0.165±0.012 &0.156±0.011 &0.527±0.006 &0.529±0.007 &0.520±0.008 &0.520±0.006\\
    &UDMF~\cite{zhang2018user} &0.702±0.016 &0.707±0.017 &0.938±0.001 &0.954±0.001 &0.963±0.002 &0.974±0.002 &0.912±0.008 &0.934±0.011\\
    &R-GCN~\cite{schlichtkrull2018modeling} &0.160±0.015 &0.158±0.014 &0.184±0.005 &0.185±0.004 &0.525±0.002 &0.536±0.003 &0.553±0.004 &0.540±0.007\\
    \cline{2-10}
    &DP-R-GCN &0.134±0.005 &\underline{0.135±0.006} &0.160±0.005 &0.157±0.007 &0.474±0.006 &0.497±0.012 &\underline{0.411±0.012} &\underline{0.417±0.009}\\
    &DP-CroSUE &\underline{0.132±0.015}  &0.136±0.017 &\underline{0.134±0.005} &\underline{0.128±0.005} &\underline{0.466±0.016} &\underline{0.474±0.017} &\textbf{0.405±0.015} &\textbf{0.397±0.013}\\
    &CroSUE &0.150±0.014 &0.155±0.012 &0.198±0.005 &0.194±0.003 &0.539±0.003 &0.543±0.002 &0.488±0.015 &0.482±0.014\\
    \toprule
    
\end{tabular}
\label{table_attack}
\end{table*} 

\subsection{User Interest Prediction (Q1)}
All models’ performances on predicting user interests are summarized in Table~\ref{table_interest}. DP-CroSUE outperforms all baselines and is only next to CroSUE, which is the same method using the unsanitized network data. Generally, single-view based embedding methods - DeepWalk, SANTEXT and Word2vec get the worst performances. Compared to single-view methods, SNE which incorporates structure and attributes together works better. It achieves maximum relative improvements of 14.1\% in precision on Twitter compared to DeepWalk. Moreover, UDMF and R-GCN which incorporate more kinds of information perform even better than SNE on most datasets. For example, compared to SNE, R-GCN gets an improvement of 13.4\% in precision on Sub-Weibo1. These observations demonstrate that incorporating more user information can help predict user interests. Compared with R-GCN, DP-R-GCN gets worse precision and Micro-$f1$ score. 
This is reasonable because DP-R-GCN extracts user information from the perturbed heterogeneous social network. These methods mentioned above are all single-network methods. Obviously, the best single-network method is R-GCN applied in real social networks. As observed, DP-CroSUE consistently outperforms R-GCN on
all the datasets. For example, the precision of DP-CroSUE is 2.0\% higher than R-GCN on Sub-Weibo1 dataset. This indicates that DP-CroSUE, though operating on perturbed networks for the purpose of information protection, can achieve satisfactory results. In addition, CroSUE achieves the best performance. This further
confirms the effectiveness of fusing networks together. However, since CroSUE works directly on real social networks, it faces user information leakage problems.

\subsection{Attribute Inference Attacks (Q2)}
Table~\ref{table_attack} shows the results of the attribute inference
attack models depicted in Section~\ref{Sec_evaluation_metric} on all the datasets. 
Note that lower scores show higher resistance to inference attacks. As shown in Table~\ref{table_attack}, all baselines operating on real social networks except for DeepWalk achieve high precision and Micro-$f1$ score in attribute inference attacks.
For example, the precision scores of DeepWalk on all datasets are nearly 50.0\%.
That means DeepWalk has no ability to infer user's gender.
\textbf{This is predictable as DeepWalk only extracts network structure information but does not involve user personal attribute features at all. } 
It is also worth noting that DP-CroSUE can achieve the second or third best results on almost all datasets. \textbf{Given that DeepWalk does not have a chance to learn user attribute features during training, the result that DP-CroSUE is only worse than DeepWalk highly validates the capability of DP-CroSUE in private attribute protection.} 
Meanwhile, the better performance of DP-R-GCN compared to R-GCN also indicates the effectiveness of our hybrid DP mechanism. For example, compared to R-GCN, the precision of DP-R-GCN drops significantly on Foursquare dataset in both gender inference and occupation inference (22.3\% and 2.6\% reduction).  
In addition, we noticed that DP-CroSUE, which incorporates information from two perturbed networks, performs even better in resisting attribute inference attacks than DP-R-GCN on most datasets. 
The reason is that if a user's attribute is disturbed into fake in one network, but remains true in another, linking these two nodes may result in both being inferred to false.
This further shows the superiority of DP-CroSUE in privacy protection.

\begin{figure*}[t]
    \centering
    \includegraphics[width=0.88\textwidth]{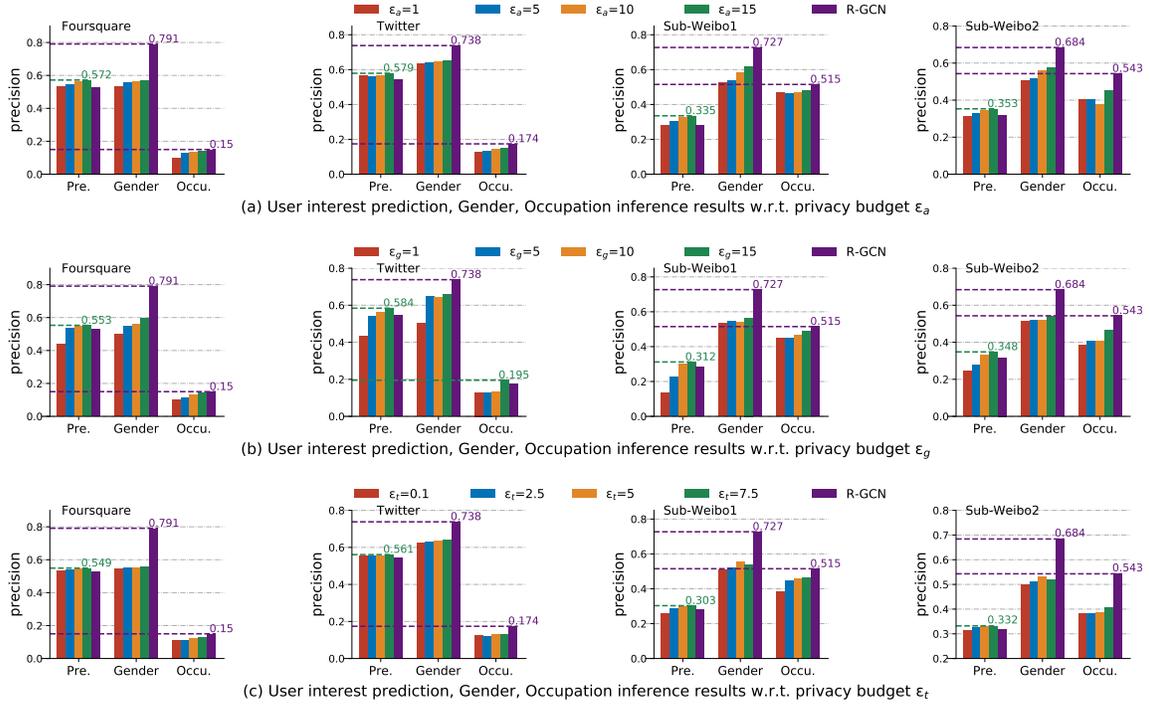}
    \vspace{-6mm}
    \caption{
    User interest prediction and privacy protection results w.r.t. privacy budget $\epsilon_a$, $\epsilon_g$, $\epsilon_t$. 
    }
    \label{Fig_perturbation}
    \vspace{-2mm}
\end{figure*}

\subsection{Accuracy and Privacy (Q3)}
To investigate how does each part of the perturbation paradigm of DP-CroSUE affect its performance on user interest prediction and privacy protection, we independently vary the privacy budget $\epsilon_a$, $\epsilon_g$ and $\epsilon_t$, and report the new results in Figure~\ref{Fig_perturbation}.

$\bullet$ \textbf{Impact of user attribute perturbation privacy budget $\epsilon_a$.}
We vary the privacy budget $\epsilon_a \in \{1, 5, 10, 15\}$. The results are shown in Figure~\ref{Fig_perturbation}(a). In general, for all the datasets, the results of user interest prediction get improvement with the increase of $\epsilon_a$. Because larger $\epsilon_a$ means less privacy thus more accurate user attributes, which helps predict user interests. For example, if we know a user is a woman, she may be more interested in fashion compared to a man.
Meanwhile, for all the $\epsilon_a$ we selected, DP-CroSUE consistently outperforms R-GCN in terms of user interest prediction. This further confirms the effectiveness of our model. As for the privacy protection capabilities, larger $\epsilon_a$, negatively influences the privacy protection results. Because more accurate user attributes are encoded into the final user representations. However, the gender and occupation information still can be effectively protected even when $\epsilon_a$ is set to a large value. From Figure~\ref{Fig_perturbation}(a), even when $\epsilon_a$ is set to 15, the precision scores of attribute inference attacks are lower than R-GCN. Our analysis is that by aligning and aggregating two perturbed datasets together, user's gender and occupation information may be further masked. Since if a user's gender or occupation is perturbed into fake in one network with the user attribute perturbation algorithm, his/her attribute in the other network may also be influenced. All the above results validate the superiority of DP-CroSUE in privacy protection.

$\bullet$ \textbf{Impact of user edge perturbation privacy budget $\epsilon_g$.}
We choose the value of $\epsilon_g$ from $\{1, 5, 10, 15\}$ and plot the results in Figure~\ref{Fig_perturbation}(b). 
Similarly, larger $\epsilon_g$ results in better user interest prediction performance.
However, the prediction precision degrades significantly when the privacy budget is small. For example, when $\epsilon_g = 1$, the precision is even lower than R-GCN. As we know, the nature of GCN is smoothing - making connected nodes similar. Real social networks have the homophily property. People tend to follow those who have similar interests with them. If we inject a large amount of noise into the network structure, the homophily level of the graph will be impacted. Thus user prediction results will decrease while the user information can be better protected. In this way, we should choose a proper $\epsilon_g$ to achieve a good trade-off between privacy protection and recommendation accuracy. 


$\bullet$ \textbf{Impact of user text perturbation privacy budget $\epsilon_t$.}
We study the impact of privacy budget in text perturbation with $\epsilon_t \in \{0.1,2.5,5,7.5\}$. As we can see from Figure~\ref{Fig_perturbation}(c), in general, larger $\epsilon_t$ brings higher user interest prediction precision. Meanwhile, we find that the improvement is relatively slow with the increase of $\epsilon_t$. The reason behind is the high utility property of the relaxed MDP notion applied in the textual data perturbation algorithm. Even when privacy budget is comparably small, with the computed substitution words, the original semantics may still be kept. Thus the user interest prediction performance can be guaranteed. What's more, it is worth noting that the perturbation of user text effectively preserves user attribute information compared to R-GCN. Since users may declare their gender and occupation in the original posts, substituting the original words can prevent information leakage effectively. For example, the word "girl" may be substituted by the word "boy". 

To sum up, to make a good trade-off of interest prediction tasks and privacy protection, we recommend a comparably small attribute privacy budget, and comparably large edge and text privacy budgets. A good budget setting can be: $\epsilon_a=5, \epsilon_g=10$ and $\epsilon_t=7.5$. The total privacy budget is $\epsilon_a + \epsilon_g +\epsilon_t\cdot d_{euc} = 15+7.5\cdot d_{euc}$.

    

\subsection{Ablation Study (Q4)}
To study the performance of the novel cross-network GCN embedding model in network integration and user interests prediction, we evaluate different
degraded model versions and record the precision scores in Table~\ref{ablation}. For convenience, we denote the original cross-network GCN model introduced in Section~\ref{Sec_crossEMB} as CroGCN. We create CroGCN-NH by replacing the hierarchy intra-graph propagation layer with iterative propagation layers like standard GCN model. Besides, we also remove the hard alignment regularization and denote the new version as CroGCN-NA. As shown in Table~\ref{ablation}, on all the datasets, CroGCN gets highest user interest prediction scores. 
The better performance of CroGCN compared with CroGCN-NH demonstrates the superiority of the hierarchy intra-graph propagation layer in transferring knowledge to the whole network. Meanwhile, the worse performance of CroGCN-NA also proves the importance of the hard alignment regularization.


\begin{table}[h]
    \caption{Ablation results of CroGCN in user interest prediction tasks.}
    \centering
    \renewcommand\arraystretch{0.95}
    \setlength{\tabcolsep}{0.8mm}
    \begin{tabular}{c|c|c|c|c}
    \toprule
    Methods &Foursquare &Twitter &Sub-Weibo1 &Sub-Weibo2 \\
    \hline
    CroGCN-NH&0.539±0.006 &0.543±0.002 &0.298±0.006 &0.325±0.009\\ 
    \hline
    CroGCN-NA&0.538±0.009 &0.545±0.004 &0.291±0.005 &0.322±0.008\\ 
    \hline
    CroGCN&0.549±0.006 &0.561±0.001 &0.303±0.003 &0.332±0.003  \\
    \toprule
    
\end{tabular}
\label{ablation}
\end{table}
\section{Conclusion}
In this work, we propose DP-CroSUE, which obtains privacy-preserving cross-network social user embeddings. DP-CroSUE perturbs users’ different information including attribute features, friendship relations and user posts through a hybrid-DP mechanism to allow further data sharing. Next, through embedding-based alignment, anchor nodes can be found and different
social networks can be learnt jointly in a pairwise manner without knowing the real user information. 
Extensive experiments demonstrate that DP-CroSUE simultaneously guards users against personal attribute inference attacks and maintain great utility in interest prediction tasks. 
Noted that DP-CroSUE can be easily generalized to other tasks like recommendation, exploitation on other tasks is left to future works.

\begin{acks}
The authors of this paper were supported by the National Key R\&D Program of China through grant 2021YFB1714800, NSFC through grant U20B2053, S\&T Program of Hebei through grant 21340301D, Beijing Natural Science Foundation through grant 4222030, and the Fundamental Research Funds for the Central Universities. 
Philip S. Yu was supported by NSF under grants III-1763325, III-1909323, III-2106758, and SaTC-1930941.
Thanks for computing infrastructure provided by Huawei MindSpore platform.
For any correspondence, please refer to Lei Jiang and Hao Peng.
\end{acks}


\bibliographystyle{ACM-Reference-Format}
\bibliography{sample-base}


\end{document}